\documentclass[12pt]{article}
\usepackage{amssymb,amsmath,epsfig}
\begin{document} \parindent=0pt
\parskip=6pt \rm

\begin{center}
{\bf \LARGE Diamagnetic critical singularity in unconventional ferromagnetic superconductors}

\vspace{0.3cm}

{\large Humberto Belich$^{1}$ and Dimo I. Uzunov$^{2\dag}$}

\end{center}

\normalsize
\vspace{0.5cm}

$^{1}$ Universidade Federal do Esp\'{\i}rito Santo (UFES),
Departamento de F\'{\i}sica e Qu\'{\i}mica, Av. Fernando Ferrari
514, Vit\'{o}ria, ES, CEP 29075-910, Brazil,

$^{2}$ Collective Phenomena Laboratory, Institute of Solid State Physics,
Bulgarian Academy of Sciences, BG-1784 Sofia, Bulgaria.

$^{\dag}$ Corresponding author. Electronic address:
d.i.uzunov@gmail.com

\vspace{0.5cm}

\begin{abstract}
The scaling properties of  the free energy, the diamagnetic moment, and the diamagnetic susceptibility above the phase transition from the ferromagnetic phase to the phase of coexistence of ferromagnetic order and superconductivity in unconventional ferromagnetic superconductors with spin-triplet (p-wave) electron paring are considered. The crossover from weak to strong magnetic induction is described for both quasi-2D (thin films) and 3D (bulk) superconductors. The singularities of diamagnetic moment and diamagnetic susceptibility are dumped for large variations of the pressure and, hence, such singularities could hardly be observed in experiments. The results are obtained within Gaussian approximation on the basis of general theory of ferromagnetic superconductors with p-wave electron pairing.

\end{abstract}
\vspace{0.5cm}


{\bf key words:}  Ginzburg-Landau theory, superconducting
fluctuations, superconductivity, ferromagnetism, magnetization.

\vspace{0.5cm}

\section{Introduction}

The discovery of coexistence of ferromagnetism and bulk superconductivity
in Uranium-based intermetallic compounds, UGe$_2$~\cite{Saxena:2000,Huxley:2001,Tateiwa:2001},
URhGe~\cite{Aoki:2001}, UCoGe~\cite{Huy:2007, Huy:2008} has led to renewed interest in
the interrelationship between ferromagnetism and superconductivity. In these itinerant
ferromagnets, the phase transition to superconductivity state occurs in the domain of
stability of ferromagnetic phase, including sub-domains, where
a considerable spontaneous ferromagnetic moment $\mathbf{M}$ is present.
This seems to be a general feature of ferromagnetic
superconductors with spin-triplet ($p$-wave) electron
pairing~\cite{Shopova:2003, Cottam:2008, Shopova:2009} (see also
reviews~\cite{Shopova:2005, Shopova:2006}). In such situation the
thermodynamic properties near the phase transition line may differ
from those known for the superconducting-to-normal metal
transition.

\begin{figure}
\includegraphics[width=8cm, height=6cm, angle=0]{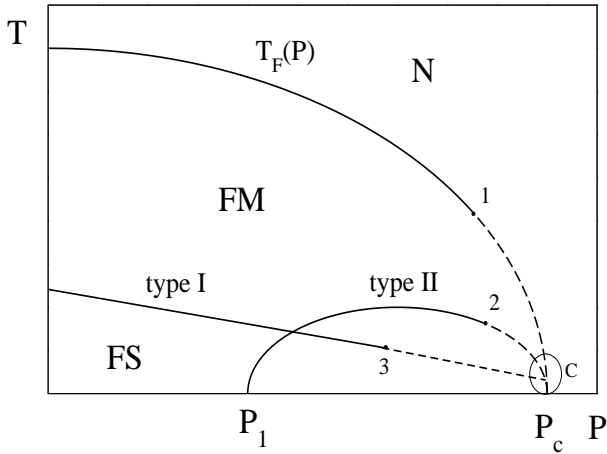}
\caption{\label{Fig1} An illustration of $T-P$ phase diagram
of $p$-wave ferromagnetic superconductors (details are omitted): N -- normal phase, FM -
ferromagnetic phase, FS - phase of coexistence of ferromagnetic
order and superconductivity, $T_{F}(P)$ and $T_{FS}(P)$ are the
respective phase transition lines: solid line corresponds to
second order phase transition, dashed lines correspond to first
order phase transition; $1$ and $2$ are tricritical points;
$P_c$ is the critical pressure, and the circle $C$ surrounds
 a relatively small domain of high pressure and
low temperature, where the phase diagram may have several forms depending on
the particular substance. The line of the FM-FS phase transition may
extend up to ambient pressure (type I ferromagnetic superconductors),
or, may terminate at $T=0$ at some high pressure $P=P_1$
(type II ferromagnetic superconductors, as indicated in the figure). }
\end{figure}

The basic thermodynamic properties of these systems are contained
in their $T-P$ phase diagrams. According to the
experiments~\cite{Saxena:2000,Huxley:2001,Tateiwa:2001, Aoki:2001,
Huy:2007, Huy:2008} carried on the above mentioned compounds and
ZrZn$_2$~\cite{Pfleiderer:2001}, the $T-P$ diagrams exhibit
several basic features, which are shown in Fig.~\ref{Fig1} (as
claimed in Ref.~\cite{Yelland:2005}, owing to a special treatment
of sample surfaces, only a surface superconductivity has been
proven in ZrZn$_2$). As seen from Fig.~\ref{Fig1}, the phase
transition line $T_{F}({P})$, corresponding to the phase
transition from normal (paramagnetic) state (N) to ferromagnetic
phase (FM) is substantially above the lines referring to phase
transition from FM phase to the phase of coexistence of
ferromagnetism and superconductivity (in short, FS phase, or, FS).
The exception is only very near to the critical pressure $P_{c}$,
where both ferromagnetism and superconductivity vanish and phase
transition lines are close to each other. This picture reflects
the real situation in the above mentioned compounds, for example,
UGe$_{2}$,  where the $T_{F}(P)$ at ambient pressure $P_a$ is of
order $53$ K, whereas the maximal $T_{FS}$ does not exceed $1.23$
K; for UGe$_2$, $P_c \sim 1.6$ GPa.

Moreover, Fig.~\ref{Fig1} shows that the line $T_{FS}(P)$ of FM-FS
phase transition may have two or more distinct shapes. Beginning
from the maximal (critical) pressure $P_c$, this line may extend
to all pressures $P < P_c$, including the ambient pressure $P_a$;
see the almost straight line containing the point 3 in
Fig.~\ref{Fig1}. A second possible form of this line, revealed by
experiments, for example, in UGe$_2$, is shown in Fig.~\ref{Fig1}
by the curve which begins at $P \sim P_c$, passes through the
point 2, and terminates at some pressure $P_{1} > P_a$, where the
superconductivity vanishes. These are two qualitatively different
physical pictures: (a) when the superconductivity survives up to
ambient pressure, and (b) when the superconducting states are
possible only at relatively high pressure (for UGe$_2$, $P_1 \sim
1$ GPa).

Within the general phenomenological theory of spin-triplet
ferromagnetic superconductors~\cite{Shopova:2003}, these different
pictures are distinguished by simple mathematical conditions on
the theory parameters~\cite{Cottam:2008, Shopova:2009}. Therefore,
there are both experimental and theoretical arguments to classify
these superconductors in ``type I'' and ``type II'' $p$-wave
ferromagnetic superconductors, as proposed for the first time in
Ref.~\cite{Cottam:2008} and illustrated in Fig.~\ref{Fig1} above
the respective $T_{FS}(P)$ lines. The tricritical points 1, 2 and
3 (see Fig.~\ref{Fig1}), at which the order of the phase
transitions changes from second order (solid lines) to first order
(dashed lines) are a quite reliable experimental fact and have
already been theoretically explained~\cite{Shopova:2003}. Firstly,
the interaction between the superconducting and magnetic
subsystems naturally generates a first order phase transition
along the high-pressure part $(P\lesssim P_c)$ of the FM-FS phase
transition. Secondly, under certain circumstances, an additional
$\mathbf{M}^6$ term in the free energy may describe the
experimentally observed first order phase transition along the
high-pressure part of the N-FM line~\cite{Shopova:2003}.

In Fig.~\ref{Fig1}, the circle $C$ denotes a narrow domain around
$P_c$ at relatively low temperatures  ($T \lesssim 300$ mK), where
the experimental data are quite few and we may not reliably
conclude about the shape of the phase transition lines in this
$T-P$ domain. It could be assumed, as in the most part of the
experimental papers, that $(T=0,P=P_c)$ is the zero temperature
point at which both lines $T_{F}(P)$ and $T_{FS}(P)$ terminate. A
second possibility is that these lines may join in a single (N-FS)
phase transition line at some point $(T\gtrsim 0,
P^{\prime}_c\lesssim P_c)$ above the absolute zero. In this second
variant, a direct N-FS phase transition occurs, although this
option exists along a very small piece of N-FS phase transition
line: from point $(0,P_c)$ to point  $(T\gtrsim 0,
P^{\prime}_c\lesssim P_c)$. A third variant is related with the
possible splitting of point $(0,P_c)$, so that the N-FM line
terminates at $(0,P_c)$, whereas the FM-FS line terminates at
another zero temperature point $(0, P_{0c})$; $P_{0c} \lesssim P_c
$. In this case, the $p$-wave ferromagnetic superconductor has
three points of quantum (zero temperature) phase
transitions~\cite{Shopova:2009}.

These and other possible shapes of  $T-P$ phase diagrams are described
within the framework of the general theory of Ginzburg-Landau (GL)
type~\cite{Shopova:2003} in an entire conformity with
the experimental data~\cite{Cottam:2008, Shopova:2009}; for reviews,
see Refs.~\cite{Shopova:2005, Shopova:2006}. The same theory has been
confirmed by a microscopic derivation based on a microscopic Hamiltonian including a
spin-generalized BCS term and an additional Heisenberg exchange term~\cite{Dahl:2007}.

Although the theory predicts correctly the shape of FM-FS phase
transition line, the possible types of phase transitions, finite
temperature and quantum multi-critical points~\cite{Shopova:2009,
Uzunov:2006}, some important features of $p$-wave ferromagnetic
superconductors, in particular, the properties of the FM-FS phase
transition, need a further investigation. The FM-FS phase
transition in $p$-wave superconductors is remarkable for the circumstance that there
the superconductivity appears in an environment of a strong
ferromagnetic moment (magnetization density) $\mathbf{M}$. This
certainly leads to a modification of the usual phase transition to
superconducting state in conventional non-magnetic
superconductors.

In this paper we study the effect of the spontaneous ferromagnetic
moment $\mathbf{M}$ deeply below the ferromagnetic transition on
the fluctuation properties in a close vicinity of the FM-FS phase
transition above the FM-FS line. Remember, that near the usual
critical point of standard (conventional) superconductors, the
fluctuating superconduction field $\psi(\mathbf{x})$ creates an
overall diamagnetic moment $V\mathbf{M}_{dia}\equiv
\mathbf{\cal{M}} = \left\{{\cal{M}}_j; j=1,2,3 \right\}$ which,
except for particular circumstances, depends on the external
magnetic field $\mathbf{H}$. The numerous studies of the
fluctuation diamagnetism in conventional superconductors,
including various sample geometries [three-dimensional (3D)
(bulk), quasi-2D (in short, q2D; thin films), 1D (wire), 0D (small
drop)] and layered structures, are summarized in
Ref.~\cite{Larkin:2005}; $D$ denotes the effective dimension of
the superconductor places in a three dimensional space ($d=3$);
note, that the effective dimension of the superconductor $D$ is
often different from the space dimensionality $d$.

Here we consider for the first time the diamagnetic properties of
3D and q2D  $p$-wave ferromagnetic superconductors --- diamagnetic
moment ${\cal{M}}$  and diamagnetic susceptibility $\chi_{dia}$.
Some of our preliminary results for 3D (bulk) superconductors have been
recently published~\cite{Belich:2010}. We demonstrate
that the diamagnetic susceptibility $\chi_{dia}$ and diamagnetic
moment ${\cal{M}}$ of $p$-wave ferromagnetic superconductors are
dumped along the prevailing part of FM-FS phase transition line, in particular,
at the second order phase transition line. So, our
theoretical results predict that the singularities at critical
points, typical for usual superconductors, do not exist at the
second order phase transition line $T_{FS}(P)$ of $p$-wave
ferromagnetic superconductors.

In Sec. II we present the GL fluctuation Hamiltonian of $p$-wave
ferromagnetic superconductors in Gaussian approximation for the
fluctuation field $\psi(\mathbf{x})$ of the superconducting order
parameter $\psi$. We consider temperature $T \geq T_{FS}(P)$,
where the statistical average $\langle\psi(\mathbf{x})\rangle$ of
$\psi =\langle\psi(\mathbf{x})\rangle + \delta\psi(\mathbf{x})$ is
equal to zero, namely, the field $\psi(\mathbf{x})$ is a net
fluctuation: $\psi \equiv \delta\psi(\mathbf{x})$. The Gaussian
approximation is the usual tool for study of basic properties of
fluctuation diamagnetism~\cite{Larkin:2005}, and here we follow
this approach as well as the notations in Refs.~\cite{Shopova:2003,
Shopova:2009, Uzunov:2010, Pitaevskii:1980}. We shall essentially
use formulations and results  of the general phenomenological
theory of $p$-wave ferromagnetic
superconductors~\cite{Shopova:2003, Cottam:2008, Shopova:2009, Shopova:2005,
Shopova:2006}, as well as modern concepts of crossover and
critical phenomena~\cite{Uzunov:2010}. Our theoretical approach essentially
generalizes preceding theoretical treatments (see, e.g., Ref.~\cite{Larkin:2005}
and references therein) and might be used as an advanced calculational scheme
in further investigations of critical crossovers in complex superconductors.

In Sec. III we derive a general expression for the fluctuation
contribution to the equilibrium free energy of system above the
curve $T_{FS}(P)$ for arbitrary values of magnetic induction
$\mathbf{B}=\mathbf{H}+ 4\pi\mathbf{M}$. We show that apart of special role of magnetic
induction $\mathbf{B}$, this expression is very similar to that
for conventional nonmagnetic superconductors.

In Sec. IV we calculate the diamagnetic moment
${\cal{M}}(T,\mathbf{H})$ and the diamagnetic susceptibility
$\chi_{dia}(T,\mathbf{H})$. Further, we investigate the ``weak-$B$
-- strong-$B$'' crossover in the behavior of these important
quantities. For this aim, both 3D and q2D sample geometries are
considered and the results are compared with known studies of
conventional ($s$-wave) non-ferromagnetic superconductors. We
predict a dumping of the singularities of the diamagnetic moment
and the diamagnetic susceptibility along the FM-FS phase
transition line. Moreover, we deduce a universality in the
behavior of these quantities in $p$-wave ferromagnetic
superconductors and usual ($s$-wave) non-magnetic superconductors.
We will also briefly focus on related experimental problems. In
Sec. IV.C and Sec. V we summarize our main results and discuss
their applicability to real systems.

\section{\label{sec:level1}Fluctuation Hamiltonian}

The fluctuation Hamiltonian of $p$-wave ferromagnetic
superconductor can be given in the form~\cite{Shopova:2003}

\begin{align}
\label{Eq1} {\cal{H}}(\psi) &= \int d^3 x
\left[\frac{\hbar^2}{4m}\sum_{j=1}^{3}\left|\left(\nabla
 - \frac{2ie}{\hbar c}\mathbf{A}\right)\psi_j\right|^2
+ a_s|\psi|^2 \right.  \nonumber \\ &
+ \frac{b_s}{2}|\psi|^4
\left. + i\gamma_0\mathbf{M}.\left(\psi\times\psi^{\ast}\right)
+\delta_0\mathbf{M}^2|\psi|^2 \right],
\end{align}
\noindent where $2m$ and $2e=2|e|$ are the effective mass and the
charge of the electron Cooper pairs, respectively, the
superconducting order parameter $\psi = \{\psi_j(\mathbf{x}); j =
1,2,3\}$ is a three-component complex field, the magnetization
$\mathbf{M}(\mathbf{x})= \left\{M_j(\mathbf{x}); j=1,2,3\right\}$,
describing the ferromagnetic order in the FM phase is a three
component real field, i.e., the components $M_j(\mathbf{x})$ are real
fields, $a_s = \alpha_s(T-T_s)$ is represented by the generic
critical temperature $T_s$ of hypothetical pure superconducting
state ($|\psi|>0, \mathbf{M}=0$) and the positive material
parameter $\alpha_s$; as usual, $b_s >0$. Besides, $\gamma_0 \sim J
>0$, where $J>0$ is an effective ferromagnetic exchange constant,
and $\delta_0$ are parameters describing the interaction betwee
superconducting and magnetic electron subsystems. As usual, the
vector potential $\mathbf{A} = \left\{A_j(\mathbf{x})\right\}$ obeys the Coulomb gauge $\nabla .\mathbf{A} = 0$ and    is related to the magnetic induction $\mathbf{B}$. In a space of dimensionality $d$, in short, $dD$ dimensional space, the relation between $\mathbf{A}$ and $\mathbf{B}$ is represented in different mathematical forms.
Here we work in a three dimensional space (d=3), where the
geometry of superconductor body is chosen of two types:  q2D (thin
film) or 3D (bulk); hence, we may use the relation $\mathbf{B} = \nabla \times \mathbf{A}$.
We neglect the gradient anisotropy~\cite{Sigrist:1991} as it has a small effect compared
to the exchange interactions between the normal and superconducting electrons; this point is discussed in the Section
below.

In mean field (MF) approximation~\cite{Uzunov:2010, Pitaevskii:1980}, the magnetization $M =
|\mathbf{M}|$ in the pure ferromagnetic phase (FM) ($\psi = 0,
\mathbf{M} \neq 0$) at zero external magnetic field ($\mathbf{H}=0$)
is given by $M_{MF}= (-a_f/b_f)^{1/2}$ as a global minimum of the
magnetic fluctuation Hamiltonian

\begin{equation}
\label{Eq2} {\cal{H}}(M) = \int d^3 x \left[c_f \sum_{j=1}^{3}\left(\nabla M_j\right)^2
+ a_f M^2 + \frac{b_f}{2}M^4\right],
\end{equation}

\noindent where $a_f = \alpha_f(T-T_F)$, $T_F$ is the generic
critical temperature of the (pure) ferromagnetic state, and
$\alpha_f$ and $b_f$ are positive material parameters~\cite{Shopova:2003}. Note, that
throughout this paper $\mathbf{M}$ denotes the density of the magnetization, whereas
the total magnetization of the ferromagnetic phase is given by $\int d^3x \mathbf{M}(\mathbf{x})$.

In the present investigation we ignore the fluctuations of the
field $\mathbf{M}(\mathbf{x})$. This is justified by the fact,
that $T_{F}(P) \gg T_{FS}$ for almost all variations of the
pressure $P$, except for a very narrow domain near the critical
pressure $P_c$, where the mean-field value $M_{MF}=
(-a_f/b_f)^{1/2}$ of the magnetization tends to zero ($T_{FS}\sim
T_{F}$). Remind, in this domain both N-FM and FM-FS phase
transitions are of first order, and therefore strong
fluctuations of $\mathbf{M}$ could not be expected.

The last two terms in Eq.~(\ref{Eq1}) have a key role in the
description of thermodynamics and phase diagram of this type of
ferromagnetic superconductors. These terms describe the
interaction between the normal (non-superconducting) electron
fraction and the electron Cooper pairs. While the $\delta_0$-term
has the supporting role of ensuring a stability of phases at
relatively large negative values of the parameter $a_f$, the
$\gamma_0$-term has a key role in the phenomenon of coexistence of
superconductivity and ferromagnetism. The presence of this term
ensures the theoretical description of real situation in $p$-wave
ferromagnetic superconductors, in particular, a reliable
description of the mentioned coexistence of phases. This is the
term which triggers the superconductivity~\cite{Shopova:2003}; for
a more detailed discussion, see Refs.~\cite{Shopova:2003,
Shopova:2005, Shopova:2006}. This point is not trivial. Although
such ''trigger'' terms, where one of ordering fields is present by
its first power and the second order parameter interacts by its
second power, are well known, for example, in the theory of
certain improper ferroelectrics~\cite{Cowley:1980}, here the
symmetry of the $\gamma_0$-term is very particular and leads to
critical phenomena of essentially new universality
class\cite{Uzunov:2006}.

Note, that according to the symmetry analysis in
Ref.~\cite{Samokhin:2002}, a $\gamma_0$-term follows from the
gradient anisotropy, which is typical for unconventional
superconductors~\cite{Sigrist:1991}. However, if the gradient
anisotropy is the only source of creating such important term, the
thermodynamics of these systems could not be described in
compliance with the experimental data. This is so, because the
exchange energy that creates the ferromagnetic order is of much
bigger magnitude than the potential corresponding to the
occurrence of FS. This argument is readily justified by the
inequalities $T_F \gg T_{FS} \gg T_s \sim 0$, which follow from
the experimental data. Therefore, the major contribution to the
$\gamma_0$-term and to the effective interaction parameter
$\gamma_0$, comes from the interaction between the Cooper pair
fraction and the normal (non-superconducting) electrons of
conduction electron bands of compounds rather than from the
exchange interaction between Cooper pairs in the superconducting
sub-band only. So, contrary to the consideration in
Ref.~\cite{Samokhin:2002}, where only the superconducting sub-band
takes part in the description and $\gamma_0$-term is a net product
of the gradient anisotropy of $p$-wave Cooper pairs, within
the present general quasi-phenomenological approach, the
interaction parameter $\gamma_0$ includes both the inter-sub-band
exchange interaction between the magnetic moments of Cooper pairs
and normal electrons, and the exchange interactions within the
sub-band of Cooper pairs, namely, between the Cooper pairs
themselves. Obviously, the second type of interaction is much
weaker, and can be safely ignored in many calculations. This point
of view is supported by recent microscopic
theories~\cite{Dahl:2007,Powell:2003}.

As known, for the Uranium-based compounds, all these remarkable
phenomena are created by the $5f$-electrons, whereas for ZrZn$_2$
the same phenomena might be ascribed to the behavior of the
$4d$-electrons. Furthermore, within our approach one may extend
the consideration beyond the itinerant ferromagnetism and take
into account exchange effects on the conduction band electrons produced by localized
spins, attached to atoms at the vertices of crystal lattice.

\section{Free energy}

Here our task is to calculate the fluctuation part of free energy

\begin{equation}
\label{Eq3} F = -\beta^{-1}\ln\int\prod_{i,j=1}^{3}\prod_{\mathbf{x} \in
V}{\cal{D}}\psi_j(\mathbf{x}){\cal{D}}\delta M_i(\mathbf{x})\exp{\left[-\beta
{\cal{H}}\right]}
\end{equation}
\noindent in the superconductor volume $V=L_xL_yL_z$, above
the FM-FS phase transition line $T_{FS}(P)\equiv T_c(P)$;
henceforth we shall use the notation $T_c$ for the temperature
$T_{FS}$. In Eq.~(\ref{Eq3}), the functional integral is over all
independent degrees of freedom: the complex components of the
superconducting fluctuations, $\psi(\mathbf{x})$ and the
fluctuations  $\delta M_i(\mathbf{x})$ of the magnetization vector
components $M_i$; $\beta^{-1} = k_BT$. The functional integral is
taken over both real [$\Re\psi(\mathbf{x})$] and imaginary
[$\Im\psi(\mathbf{x})$] parts of the complex field
$\psi(\mathbf{x})$, i.e., ${\cal{D}}\psi(\mathbf{x}) \equiv
d\Re\psi(\mathbf{x})d\Im\psi(\mathbf{x})$. Note that for
temperatures near $T_{c}(P)$ we can always set $\beta
\approx\beta_{c}= 1/k_BT_{c}$ (see, e.g.,
Ref.~\cite{Uzunov:2010}). Mostly in this paper we shall consider
3D (bulk) superconductors and (q2D) thin superconducting films in
a transverse magnetic field, so we shall use the 3D notations:
$\mathbf{x}=(x,y,z)$, and the labels $x,y,z$ for quantities
defined along the respective Cartesian axes in 3D space ($d=3$).

We are interested in the magnetic thermodynamics in close vicinity
($T\sim T_{c}$) above the temperature $T_{c}(P)$ of FM-FS phase
transition. For $T_F \gg T_{c}$, the fluctuations $\delta M_i$ in
this temperature range are very weak and can be neglected [$M_i
\approx M_i^{\scriptsize (MF)}$]. Then one may substitute $
M_j(x)$ in the exponent of Eq.~(\ref{Eq3}) by
$M_j^{\scriptsize(MF)}$ and ${\cal{D}} \delta M_j(x)$ in the
functional integration by ${\cal{D}}\delta[\delta M_j(x)]$; here
$\delta[z]$ denotes $\delta$-function. This procedure totally
eliminates the fluctuations $\delta {M}_{i}(\mathbf{x})$ from our
consideration; hereafter we shall omit the label ``MF''
[$\mathbf{M}^{\scriptsize (MF)} \equiv \mathbf{M}$]. For
$T>T_{FS}$ the statistical averages $\langle\psi_j\rangle = \psi_j
- \delta \psi_j$ are equal to zero and we have a purely
fluctuation field $\psi(\mathbf{x}) = \delta\psi(\mathbf{x})$.
Thus, within the present consideration, the only integration
variables in functional integral (\ref{Eq3}) are the fluctuations
$\delta \psi_j(\mathbf{x}) = \psi_j(\mathbf{x})$. Neglecting the
Ginzburg critical region~\cite{Uzunov:2010, Pitaevskii:1980},
which is very small in all low temperature superconductors and,
hence, unobservable in experiments, we may ignore the fourth order
fluctuation term $|\psi_j|^4$ in Eq.~(\ref{Eq1}), and apply
Gaussian approximation to the fluctuation modes
$\psi(\mathbf{x})$.

For convenience, we choose the vectors $\mathbf{M}$ and
$\mathbf{H}$ along the $\hat{z}$-axis: $\mathbf{M}=(0,0,M)$, and
$\mathbf{H}=(0,0,H).$ This assumption does not essentially
restrict the generality of our consideration. If the
superconductors is magnetically isotropic, the magnetization
vector $\mathbf{M}$ will follow the direction of external magnetic
field $\mathbf{H}$. If magnetic anisotropy is present, for
example, an easy axis of magnetization, as in UGe$_2$, our
assumption will be satisfied by choosing an external field
$\mathbf{H}$ parallel to this easy axis of magnetization. Then the
term $\mathbf{M}.(\psi\times\psi^{\ast})$ takes the simple form
$M(\psi_1\psi_2^{\ast} - c.c.)$.

Under the supposition of uniform
magnetic induction $\mathbf{B} = (0,0,B)$, we take the gauge of
the vector potential $\mathbf{A}$ as $\mathbf{A} = (-By,0,0)$
 and expand the fields $\psi_j(\mathbf{x})$
in series~\cite{Pitaevskii:1980}

\begin{equation}
\label{Eq4} \psi_j(\mathbf{x}) =\sum_{q}c_j(q)\varphi_j(q, \mathbf{x})\,
\end{equation}
\noindent in terms of the complete set of eigenfunctions

\begin{equation}
\label{Eq5} \varphi_j(q, \mathbf{x}) =
\frac{1}{\left(L_xL_z\right)^{1/2}}e^{i(k_x+k_z)}\chi_n(y)
\end{equation}
\noindent of the operator $\left[i\hbar\nabla +
(2e/c)\mathbf{A}\right]^2/4m$, corresponding to the eigenvalues

\begin{equation}
\label{Eq6} E(q) =  \left(n + \frac{1}{2}\right)\hbar\omega_c +
\frac{\hbar^{2}}{4m}k_z^2,
\end{equation}
specified by the magnetic frequency $\omega_B = (eB/mc)$ and
vector $q = (n,k_x,k_z)$, where $n= 0, 1, \dots, \infty$, is the
quantum number corresponding to the Landau levels, and $k_x$ and
$k_z$ are components of the wave vector $\mathbf{k} =
(k_x,k_y,k_z)$. In Eq.~(\ref{Eq5}), the function $\chi_n(y)$ is
related to the Hermite polynomials $H_n(y)$ by

\begin{equation}
\label{Eq7}\chi_n(y) = A_n
e^{-\frac{(y-y_0)^2}{2a_B^2}}H_n\left(\dfrac{y-y_0}{a_B}\right),
\end{equation}
\noindent where $A_n^{-1} =
(a_B2^nn!\sqrt{\pi})^{1/2}$~\cite{Abramowitz:1965}, $y_0 =
a_B^2k_x$, and $a_B = (\hbar c/2eB)^{1/2}$.

In terms of the $c_j(q)$-functions, the $\psi^2$-part of the
fluctuation Hamiltonian (\ref{Eq1}) is given by

\begin{equation}
\label{Eq8} {\cal{H}}  = \sum_{j,q} \tilde{E}(q)
c_j(q)c_j^{\ast}(q) + i\gamma_0M \left[
c_1(q)c_2^{\ast}(q) - \mbox{c.c.}\right],
\end{equation}
\noindent where

\begin{equation}
\label{Eq9}\tilde{E}(q) = E(q)+a_s+\delta_0 M^2.
\end{equation}

\noindent Applying the unitary transformation,

\begin{subequations}
\begin{equation}
\label{Eq10a} c_1(q) = \frac{i}{\sqrt{2}}\left[-\phi_{+}(q) +
\phi_{-}(q)\right]
\end{equation}
\begin{equation}
\label{Eq10b} c_2(q) = \frac{1}{\sqrt{2}}\left[\phi_{+}(q) +
\phi_{-}(q)\right]
\end{equation}
\end{subequations}
\noindent renders the fluctuation Hamiltonian (\ref{Eq1}) as a sum
of squares of field components $c_3(q)$, and $\phi_{\pm}(q)$.

In the continuum limit, $L_y \rightarrow \infty$, the integral

\begin{equation}
\label{Eq11} I_y = \int_{-L_y/2}^{L_y/2}dy \chi_n(y)\chi_{n^{\prime}}(y)
\end{equation}
\noindent is simply equal to the Kronecker symbol
$\delta_{nn^{\prime}}$ and this is a key point in the further
simple representation of the Hamiltonian (\ref{Eq1}). Substituting
the function (\ref{Eq7}) in the integral (\ref{Eq11}), and having
in mind the properties of the Hermite polynomials
$H_n(z)$~\cite{Abramowitz:1965}, we see that the integral
(\ref{Eq11}) will be equal to $\delta_{n,n^{\prime}}$ only if the
limits of integration can be expanded to $\pm\infty$. In fact, the
integration in Eq.~(\ref{Eq11}) can be performed with respect to
the variable $\bar{y} = (y - y_0)/a_B$ and integral limits $( -y_0
\pm L_y/2)/a_B$. The limits of integration with respect to this
variable can be approximately equalized to
 $\pm \infty$, if only $L_y/2a_B \rightarrow \infty$; for
finite samples ($L_y \gg 2a_B$). Another inevitable condition, namely,
$-L_y/2 < y_0 < L_y/2$, follows from the requirement that
the coordinate $y_0 = a_B^2k_x$ must belong to the sample volume. This condition implies

\begin{equation}
\label{Eq12} -\frac{L_y}{2a_B^2} < k_x < \frac{L_y}{2a_B^2}.
\end{equation}

As we show below, this condition fixes the number of states
${\cal{N}} = L_xL_y/2\pi a_B^2$ for all possible values of quantum
number $k_x$ at any given $n$ and $k_z$.

Having in mind these features of theory and considering
sufficiently large $L_y$ we can justify the solution $I_y =
\delta_{nn^{\prime}}$ of the integral (\ref{Eq11}) and achieve a
very useful form of the Hamiltonian, namely,

\begin{align}
\label{Eq13} {\cal{H}} = & \sum_{n,q}\left[ E_{-}(q)|\phi_{+}(q)|^2+
E_{+}(q)|\phi_{-}(q)|^2 \right.  \nonumber \\ & \left.   + E_3(q)|c_3(q)|^2\right],
\end{align}
with
\begin{equation}
\label{Eq14} E_{\pm}(q) = E (q) + a_{\pm}(M),
\end{equation}
\noindent
where $E(n,k_z)$ is given by Eq.~(\ref{Eq6}),

\begin{equation}
\label{Eq15} a_{\pm}(M) = a_0  \pm \gamma_0 M
\end{equation}

\noindent is represented by $a_0 = a_s + \delta_0 M^2$, and $E_3 \equiv \tilde{E}$ is
given by Eq.~(\ref{Eq9}).

Now the free energy (\ref{Eq3}) can be written as a functional
integral over all independent field amplitudes: $\pm \phi(q)$, and $c_3(q)$. Using the short
notations $\varphi_{\alpha}(q)$ with $\alpha = (+,-,3)$ of the
Fourier amplitudes
$\phi_{+}(q)$, $\phi_{-}(q)$ and $c_3(q)$, respectively, and adopting
 the same label $\alpha$ to denote $E_{+}$,
$E_{-}$ and $E_{3}$ by $E_{\alpha}$, we obtain the free energy $F$ in the form

\begin{equation}
\label{Eq16} F = - \int
\prod_{\alpha,q}d\Re\varphi_{\alpha}(q)d
\Im\varphi_{\alpha}(q)e^{-\beta\sum_{\alpha,q}E_{\alpha}(q)|\varphi_{\alpha}(q)|^2}.
\end{equation}
The direct calculation of the Gaussian integrals in Eq.~(\ref{Eq16}) yields

\begin{align}
\label{Eq17} F & =
-k_BT\ln\prod_{\alpha,n,k_x,k_z}\left[\frac{\pi
k_BT}{E_{\alpha}(n,k_z)}\right]  \nonumber \\ &
 =  -k_BT\ln\prod_{\alpha,n,
k_z}\left[\frac{\pi k_BT}{E_{\alpha}(n,k_z)}\right]^{{\cal{N}}},
\end{align}
where we have used the condition (\ref{Eq12}) and the continuum limit for the $k_x$-product,

\begin{equation}
\label{Eq18}
\prod_{k_x}1 \;\;\; \longrightarrow \;\;\; \exp\left[L_x\int dk_x/2\pi\right],
\end{equation}
\noindent namely,
\begin{equation}
\label{Eq19} {\cal{N}} = \sum_{k_x} 1 \approx L_x
\int^{L_y/2a_B^2}_{-L_y/2a_B^2}\frac{dk_x}{2\pi} =
\frac{L_yL_x}{2\pi a_B^2}.
\end{equation}
\noindent In the second equality (\ref{Eq17}) we point out the
result of the summation over $k_x$ with the help of the rule
(\ref{Eq18}). The obtained expression (\ref{Eq17}) follows from
the fact that the mode energies $E_{\alpha}(n,k_z)$ do not depend
on the quantum number $k_x$. Thus the number of states ${\cal{N}}$
at fixed quantum numbers $n$ and $k_z$ and the relation $F \sim
{\cal{N}}$ are naturally deduced from the calculation.

Further, we have to pay attention to the fact that the field
theories of GL type are limited to length scales $k=|\mathbf{k}|
\lesssim \Lambda \sim \pi/\xi_0$, where $\xi_0$ is the zero
temperature correlation length of the field of
interest~\cite{Uzunov:2010, Pitaevskii:1980}. In the present case,
we must use the quantity $\xi_0$ corresponding to the field
$\psi(q)$ which fluctuates in the vicinity of the phase transition
line $T_{c}(P)$. The standard expression $\xi_{0s} =
\hbar/(4m\alpha_sT_s)^{1/2}$, corresponding to the generic
critical temperature $T_s$~\cite{Pitaevskii:1980, Uzunov:2010}
cannot be applied to our problem. So, we define the upper cutoff
$\Lambda \simeq \pi/\xi_0$ by the zero-temperature correlation
(coherence) length $\xi_0$ but the latter will be specified at a
next stage of our consideration. Here we will mention that $\xi_0$
is the scaling amplitude of the correlation length $\xi$ of the
superconducting fluctuations at the FM-FS critical line
$T_{c}(P)$: $\xi(t) = \xi_0/|t|^{1/2}$, where $t =
(T-T_{c})/T_{c}$; $|t| < 1$.

As the small wave numbers $k$ have the main contribution to the
values of the integrals in the free energy and its derivatives, we
shall use the finite cutoff $\Lambda$ only when the respective
integral has an ''ultraviolet'' divergency; for example, see
Eq.~(\ref{Eq19}). In all other cases, the relatively large values
of $k$ do not produce essential quantitative contributions to the
integrals and for this reason, we may extend the cutoff $\Lambda$
to infinity. Moreover, owing to the same type of limitation of the
GL theory --- the long wavelength approximation $\xi_0k \lesssim
\pi$, we should take in mind that only quantum numbers $n$
corresponding to energies $E_{\alpha} \lesssim \hbar^2/4m\xi_0^2$
are to be taken into account. Thus the quantum number $n$ has a
cutoff as well, and the latter is given by $n_c\omega_B \simeq
\hbar/4m\xi_0^2$, namely, $n_c = [\hbar/4m\xi_0^2\omega_B]$ ($[z]$
denotes the integer part of number $z$). This energy cutoff could
be neglected in cases when this does not produce divergencies of
the respective physical quantities.

For our further aims we shall write Eq.~(\ref{Eq17}) in a more convenient form:

\begin{equation}
\label{Eq20} F  = -S\frac{e k_B T B}{\pi\hbar c} \sum_{n=0}^{n_c}
\sum_{k_z=-\Lambda}^{\Lambda}\ln\frac{(\pi k_B T)^3}{E_{+}E_{-}E_{3}} ,
\end{equation}

The relevant part of the free energy, which contains singularities
at the critical temperature $T_{c}$ is given by the term
containing $\ln E_{-}$. All other terms are quite smooth near the
line $T_{c}(P)$ and do not produce singularities of the physical
quantities. This important circumstance follows directly from the
fact, that namely the parameter $a_{-}(M)$ is relatively small in
magnitude and changes sign at $T_{c}$ --- the FM-FS phase
transition temperature, corresponding to the phase transition from
FM phase to the phase of coexistence (FS) of ferromagnetism and a
homogeneous (Meissner) superconducting state at zero external
magnetic field ($H=0$). Therefore, the critical fluctuations are
described by the field $\phi_{-}(q)$. The other fields,
$\phi_{+}(q)$ and $c_3(q)$, do not produce critical phenomena
(singularities) because the parameters $a_{+}$ and $a_0$ do not
pass through the null at $T=T_{c}$. The value of magnetization $M$
is relatively large along the most part of the line $T_{c}(P)$ in
this type of ferromagnetic superconductors $(T_F \gg T_{c})$, and
therefore the parameters  $a_{+}$ and $a_0$ are quite different
from $a_{-}$ except for a narrow domain around the critical
pressure $P_c$. Note, that in a non-magnetic (standard) $p$-wave
superconductor~\cite{Uzunov:2010}, where $M \equiv 0$,
$a_{\pm}=a_0=a_s$, all three modes $\varphi_{\alpha}(q)$ are
critical in a close vicinity of the critical point $T_s$. For such
superconductor, the free energy (\ref{Eq21}) will differ with a
factor 3 from the standard result for a conventional ($s$-wave)
superconductor with a scalar order parameter
\cite{Pitaevskii:1980}.

In our further analysis we shall ignore the nonsingular part of
the free energy and keep only the contributions from the critical
mode $\phi_{-}(q)$. Thus we have to analyze the behavior of
function

\begin{equation}
\label{Eq21} F  = -S\frac{e k_B T B}{\pi\hbar c} \sum_{n=0}^{n_c}\sum_{k_z=-
\Lambda}^{\Lambda}\ln\frac{\pi k_B T}{E(q) + a_{-}(M)},
\end{equation}
\noindent where $E(q)$ is given by Eq.~(\ref{Eq6}).

Now we have to define the parameters in Eq.~(\ref{Eq21}).
For $M = (|a_f|/b_f)^{1/2}$, the parameters $a_{\pm}(M)$ are given by~\cite{Cottam:2008, Shopova:2009}

\begin{equation}
\label{Eq22} a_{\pm}(T) = \alpha_s(T-T_s) + \delta_0\frac{a_f}{b_f}
\pm\gamma_0\left(\frac{a_f}{b_f}\right)^{1/2}.
\end{equation}
We are interested mainly on the parameter $a_{-}(T)$ which is related with the equilibrium phase
transition from FM to FS. Defining $T_c\equiv T_{FS}$ from the equation $a_{-}(T_{c}) = 0$, we obtain

\begin{equation}
\label{Eq23} a_{-}(T) \approx \alpha_c(T-T_c),
\end{equation}
where
\begin{equation}
\label{Eq24} \alpha_c = \alpha_s - \frac{\delta_0\alpha_f}{b_f} +
\frac{\gamma_0\alpha_f^{1/2}}{2\left[b_f\left(T_{F}-T_{c}\right)\right]^{1/2}}
\end{equation}
and $T_c$ is given as a solution of the equation

\begin{equation}
\label{Eq25} T_{c}(M_c) = T_s - \frac{\delta_0}{\alpha_s}M_c^2  +
\frac{\gamma_0}{\alpha_s}M_c,
\end{equation}

\noindent where $M_c \equiv
M(T_{c})=\left[\alpha_f(T_F-T_{c})/b_f\right]^{1/2} > 0$. In the
same way one obtains that the parameters $a_{0}$ and $a_{+}$
remain positive at $T_{c}$: $a_{0}(T_{c}) =\gamma_0M_c$, and
$a_{+}(T_{c})=2a_{0}(T_{c})$, which is a demonstration that the
modes $\phi_{+}(q)$ and $c_3(q)$ are not critical and could not
have essential contributions to the thermodynamics in the vicinity
of phase transition line $T_{c}(P)$.

Note that the parameter $\alpha_c$, given by Eq.~(\ref{Eq24}), is
positive for requirements of stability of the ordered
phases~\cite{Shopova:2005}. The solution of Eq.~(\ref{Eq25}) with
respect to $T_c$ yields the curve $T_{FS}(P) \equiv T_c(P)$, shown
in Fig.~\ref{Fig1}. The dependence of $T_c$ on the pressure
$P$ comes from the $P$-dependence of material parameters
($\alpha_s, \alpha_f, T_F, \dots$) in Eq.~(\ref{Eq25}). In
Refs.~\cite{Cottam:2008, Shopova:2009} a simple $P$-dependence of
these parameters has bee suggested: all material parameters except
$T_F$ are $P$-independent, and the form of the function $T_F(P)$
is assumed of the simple form $T_F(P) \approx T_{F}(0)(1 -
P/P_c)$. Although this is a simple approximation of the pressure
effect in these systems, it gives a remarkable agreement between
theory and experimental data for the $T-P$ phase
diagram~\cite{Cottam:2008, Shopova:2009}. In the framework of the
same approximation, according to Eq.~(\ref{Eq25}), the
$P$-dependence is contained in the quantity $M_c$ and, hence, we
may often consider $T_c$ as a function of $M_c \equiv M(T_c)$ --
the value of the magnetization on the FM-FS phase transition line:
$ T_c = T_c(M_c)$.

The upper cutoff for the wave number $k_z$ is given by $\Lambda
\simeq \pi/\xi_0$, where the zero temperature correlation length
$\xi_0 = (\hbar^2/4m\alpha_cT_c)^{1/2}$ is expressed by $\alpha_c$
and $T_c$, given by Eqs.~(\ref{Eq24}) and (\ref{Eq25}),
respectively. The upper quantum number $n_c$, defined by the
equality $n_c\hbar\omega_c \simeq \hbar^2/4m\xi_0^2$, can be
represented by $n_c = [1/2b]$, where $b = (e\hbar  B/2mc\alpha_c
T_c)$ is a non-negative quantity. Having in mind the supplementary
condition that  $E(n,k_z)$ from Eq.~(\ref{Eq6}) with $n=k_z=0$
should also obey the condition $E(0,0) \leq \hbar^2/4m\xi_0^2$, we
find that $b \in [0,1]$. For type II superconductors, the
parameter $b$ has the useful representation $b = B/B_{c2}(0)$ by
the upper critical induction $B_{c2}(T) = B_{c2}(0)|t|$ at zero
temperature, $B_{c2}(0) = e\hbar/2mc\alpha_cT_c$, i.e., for $|t| =
1$; henceforth we shall denote $B_{c2}(0)$ by $B_0$.

Using these remarks, we can represent the free energy in the form

\begin{equation}
\label{Eq26} F  = \rho S B_0f(b,t)
\end{equation}
\noindent where $\rho = (e k_B T /\pi\hbar c)$, and

\begin{equation}
\label{Eq27} f(t,b) =\sum_{k_z=-\Lambda}^{\Lambda}S(k_z, t,b)
\end{equation}
is given by the sum
\begin{equation}
\label{Eq28} S(k_z,t,b)  = -b\sum_{n=0}^{[1/2b]}\ln
\dfrac{\left(\pi k_B/\alpha_c\right)\left(1+t\right)}{2bn + b + t +
\xi_0^2 k_z^2}.
\end{equation}

\noindent  The function $f(t,b)$ describes the shape of the
free energy $F(T,B)$, whereas
the functions $m(t,b) = -\partial f/\partial b$ and
$m^{\prime}(t,b) = \partial m/\partial b $ represent the variations of the diamagnetic
moment ${\cal{M}} = -\partial F(T,H)/\partial H$ and the diamagnetic
susceptibility $\chi_{dia} = -\partial^2 F(T,H)/\partial H^2$,
respectively. For the choice $\mathbf{H} = (0,0,H)$, the diamagnetic vector
$\mathbf{\cal{M}}$ has only
one component: ($0,0,{\cal{M}}$). As $\partial H = \partial B$, we can use the
formulae ${\cal{M}} = - \partial F(T,B)/\partial B$ and
$\chi_{dia} = \partial M(T,B)/\partial B$.

The wave number $k_z$ lies in the reduced Brillouin zone:
$-\pi/\xi_0 < k_z=2\pi l/L_z \leq \pi/\xi_0$; $l = 0,\pm 1, ...\pm [L_z/\xi_0]$.
For q2D systems, where $L_z \leq \xi_0$, the only
possible value of $k_z$ is zero and,
hence, for q2D superconductors the function $f(t,b)$
coincides with $S(k_z,t,b) = S(0,t,b)$. For 3D systems,
we shall use the continuum limit for $f(t,b)$, given by

\begin{equation}
\label{Eq29} f(t,b) =L_z\int_{-\Lambda}^{\Lambda}\frac{dk_z}{2\pi}S(k_z, t,b).
\end{equation}

In Eqs.~(\ref{Eq21}) and (\ref{Eq28}), the logarithmic divergence
at maximal temperature corresponds to $n=0$, $k_z=0$, and
$\epsilon = (b + t) = 0$. The parameter $t$ indicates the vicinity
to $T_c$ along the $T$-axis and the parameter $b= B/B_0$ shows the
strength of the induction $B$ and the distance to the phase
transition point $(T_c,B=M)$ along the $H$-axis of the $(T,P,H)$
phase diagram. These two parameters, $t$ and $b$, are suitable for
investigations of the system properties for $t > 0$. It is easy to
show that for $t < 0$, where the upper critical induction
$B_{c2}(T) = B_0(-t) \geq 0$ of type II superconductors is
defined, the parameter $\epsilon = (t+b)$ can be represented in
the suitable form $\epsilon= [B-B_{c2}(T)]/B_0$, or,
alternatively, in the form $\epsilon = [T-T_{c2}(B)]/T_c$, where
$T_{c2}(B) = T_c(1-b)$ is the higher critical temperature of type
II superconductors~\cite{Pitaevskii:1980}. Thus, for $t <0$, the
parameter $\epsilon$ shows the distance from the phase transition
line described by the upper critical induction $B_{c2}(T) =
B_0|t|$, or, alternatively, by the higher critical temperature
$T_{c2}(B)$. The parameter $\epsilon$ is appropriate for
investigations at temperatures $T < T_c(P)$, i.e., $t < 0$. In
this paper our consideration is restricted to temperatures $T >
T_c(P)$ and for this reason we shall use the original parameters
variables $t$ and $b$, as given in Eqs.~(\ref{Eq26}) --
(\ref{Eq29}).

Performing the summation in Eq.~(\ref{Eq28}) and keeping only
terms which depend on $b \sim B$, we obtain

\begin{align}
\label{Eq30} S(k_z,t,b)  &= -b\ln\left(\frac{\pi k_B}{\alpha_c}\right)
- b\ln(1+t) \nonumber \\ &
+ \frac{\ln b}{2} +  b\ln \left[1 + b + \varepsilon(k_z)\right]
\nonumber \\ &
+ b\ln \frac{\Gamma\left[\dfrac{1 + b + \varepsilon(k_z)}{2b}\right]}
{\Gamma\left[\dfrac{b + \varepsilon(k_z)}{2b}\right]},
\end{align}

\noindent where $\varepsilon(k_z) = \left(t + \xi_0^2k_z^2\right)$
and $\Gamma (z)$ is the gamma function. The sum $S(k_z,t,b)$,
given by Eq.~(\ref{Eq30}) and the shape function $f(t,b)$, given
by Eq.~(\ref{Eq27}), contain redundant terms. Remind that we have
neglected the contributions from the factors $E_{+}$ and $E_3$ in
Eq.~(\ref{Eq20}) although most of them depend on the parameter
$b$, namely, on the induction $B$ and, hence, these terms may have
a finite contribution to the diamagnetic moment ${\cal{M}}$. The
mentioned terms have been however ignored for the fact that they
do not produce singularities in the free energy derivatives. Thus,
within the approximations already made, we cannot evaluate
correctly the magnitude of finite contributions to important
quantities as ${\cal{M}}$ and $\chi_{dia}$ at the phase transition
point. We may just demonstrate that such contributions exist. To
be in a consistency with our preceding consideration, we should
neglect such terms in Eqs.~(\ref{Eq26}), (\ref{Eq27}) and
(\ref{Eq30}), too.

Up to now we keep all $b$-dependent terms contained in the general
formula (\ref{Eq21}) for the free energy, including  terms which
obviously does not lead to any singularities, for example, the
first and second terms on the r.h.s. of Eq.~(\ref{Eq30}). At this
stage we make the stipulation to keep these terms in our further
consideration with the remark that contributions to ${\cal{M}}$
and $\chi_{dia}$ which are finite at the phase transition point
will be neglected, provided divergent term is present. When no
divergency occurs in some of this quantities, we shall keep the
finite term only to indicate the lack of divergencies and to show
that the respective quantity remains finite at the phase
transition point.

\section{Crossover from weak to strong magnetic induction}

The free energy, the diamagnetic moment ${\cal{M}}$ and the
diamagnetic susceptibility $\chi_{dia}$ can be investigated
analytically in the limiting cases of strong and weak magnetic
induction $B$. Above the critical temperature $T_{c}$, when $t
>0$, the weak-$B$ limit is defined by the condition

\begin{subequations}
\begin{equation}
\label{Eq31a} \varepsilon(k_z) = t + \xi_0^2k_z^2 \gg b,
\end{equation}
whereas the strong-$B$ limit is given by the opposite condition

\begin{equation}
\label{Eq31b} \varepsilon (k_z) \ll b.
\end{equation}
\end{subequations}
These conditions are considered in a close vicinity ($t \ll 1$) of
the phase transition line $T_c(P)$. The condition (\ref{Eq31a}) is
satisfied for any $k_z$, provided $t \gg b$. The condition
(\ref{Eq31b}), however, can be satisfied only for $t \ll b$ and
$(\xi_0k_z)^2$ sufficiently small. For large $k_z \sim \pi/\xi_0$
this condition does not hold for any $b \in [0,1]$. However, the
sum (\ref{Eq27}) is practically taken for $k_z \neq 0$ only for
the 3D geometry, and in this case, the main contribution to the
sum is given by the relatively small wave numbers ($\xi_0k_z \ll
\pi$). Having in mind this argument, we can use the condition
(\ref{Eq31b}) without any restriction to small values of $\xi_0
k_z$ because the final result for the free energy will not
essentially depend on the contribution of relatively large wave
numbers. In particular, this is true in the continuum
limit~(\ref{Eq29}) for the sum (\ref{Eq27}) and $(t + b)\ll 1$,
namely, in the close vicinity of the phase transition line
$T_c(P)$.

 \subsection{Weak-$B$ limit}

 Applying the condition
 (\ref{Eq31a}) to the sum (\ref{Eq30}) we obtain the result

\begin{align}
\label{Eq32} S(k_z, t, b) &= b\ln\frac{1 + \varepsilon(k_z)}{(\pi
k_B/\alpha_c)(1+t)} \nonumber \\ & + \frac{b^2}{12}\left[
\frac{11}{1+\varepsilon(k_z)} + \frac{1}{\varepsilon(k_z)}\right].
\end{align}
\noindent In Eq.~(\ref{Eq32}), all $b$-independent terms have been
omitted as irrelevant to our consideration and small terms of type
$O(b^3)$ have been neglected. From Eq.~(\ref{Eq32}) is readily
seen that only the term of type $b^2/\varepsilon (k_z)$ will
produce a singularity of the thermodynamic functions as this term
tends to infinity for $k_z \sim 0$ and $t \sim 0$. Therefore, in
this case, we can neglect the other terms in Eq.~(\ref{Eq32}) and
write the free energy in the form

\begin{equation}
\label{Eq33} F  = \frac{\rho S B^2}
{12 B_{0}}\sum_{k_z=-\Lambda}^{\Lambda}\frac{1}{\varepsilon(k_z)}.
\end{equation}
This result has been obtained in Ref.~\cite{Belich:2010} in
different notations.

\subsubsection{3D superconductors}

For 3D superconductors and $t \ll 1$, the sum~(\ref{Eq27}) over
$k_z$ can be substituted by the integral~(\ref{Eq29}) and the
cutoff $\Lambda$ can be extended to infinity. Then the free energy
becomes
\begin{equation} \label{Eq34} F_{3D}  = \frac{\rho V
B^2}{24B_0\xi_0t^{1/2}}.
\end{equation}
\noindent The diamagnetic
moment ${\cal{M}}\equiv M_{dia} = -\partial F(T,B)/\partial B$ takes the form
\begin{equation}
\label{Eq35} {\cal{M}}_{3D}(T,H)  = -\frac{\rho
VB}{12B_0\xi_0t^{1/2}}.
\end{equation}
In contrast to the usual case of non-magnetic
superconductors~\cite{Larkin:2005}, here the diamagnetic moment
does not vanish at $H=0$ but rather remains proportional to the
magnetization $M_c$ at the FM-FS phase transition line. In fact,
as we work at a close vicinity of FM-FS phase transition line
$T_c(P)$, the magnetic induction in Eq.~(\ref{Eq35}) should be
approximated by $ B\approx B_c = H + 4\pi M_c $, where $M_c =
M(T_c)$. Within the weak-$B$ limit, this result is valid for
relatively small values of $M_c$: $M_c \ll B_0$.

The diamagnetic susceptibility $\chi_{dia}^{(3D)} = \partial
{\cal{M}}_{\scriptsize 3D}/\partial B$ is given by

\begin{equation}
\label{Eq36} \chi_{dia}^{(3D)}(T)  = -\frac{\rho
V}{12B_0\xi_0t^{1/2}}.
\end{equation}

In these general notations this is the well known result for the
fluctuation diamagnetic susceptibility above the critical point
$T_c$ of conventional superconductors~\cite{Larkin:2005, Pitaevskii:1980}. For
$p$-wave ferromagnetic superconductors we have to take into
account that $T_c=T_{FS}$ as given by Eq.~(\ref{Eq25}) and the
material parameter $\alpha_c$, which enters in the zero-temperature correlation length $\xi_0$,
is given by Eq.~(\ref{Eq24}). Thus one reveals the result for
$\chi_{dia}(T)$, obtained in Ref.~\cite{Belich:2010}.

\subsubsection{Quasi-2D superconductors}

For thin films, where $L_z < \xi_0 = \pi/\Lambda$, only the wave
number $k_z = 0$ satisfies the condition $ -\pi/\xi_0 < k_z = 2\pi
l/L_z \leq \pi/\xi_0 $; $l = 0, \pm 1, \dots, [\pi L_z/\xi_0]$.
For such q2D geometry, $f(t,b) = S(0,b,t)$. In the weak-$B$ limit
(\ref{Eq31a}), $S(0,b,t)$ is obtained by setting $k_z=0$ in
Eq.~(\ref{Eq32}). Once again we may keep only the leading singular
term $b^2/12t$. Thus we obtain the free energy in the form

\begin{equation}
\label{Eq37} F_{2D}  = \frac{\rho SB^2}{12B_0t}.
\end{equation}
\noindent In Eq.~(\ref{Eq37}) and below we use the label ``2D'' to denote
quantities corresponding to q2D systems.
Now one easily finds that

\begin{equation}
\label{Eq38} {\cal{M}}_{2D}(T,H)  = - \frac{\rho SB}{6B_0t}.
\end{equation}

\begin{equation}
\label{Eq39} \chi_{dia}^{(2D)}(T)  = - \frac{\rho S}{6B_0t}.
\end{equation}

\begin{figure}
\includegraphics[width=8.5cm, height=8cm, angle=0]{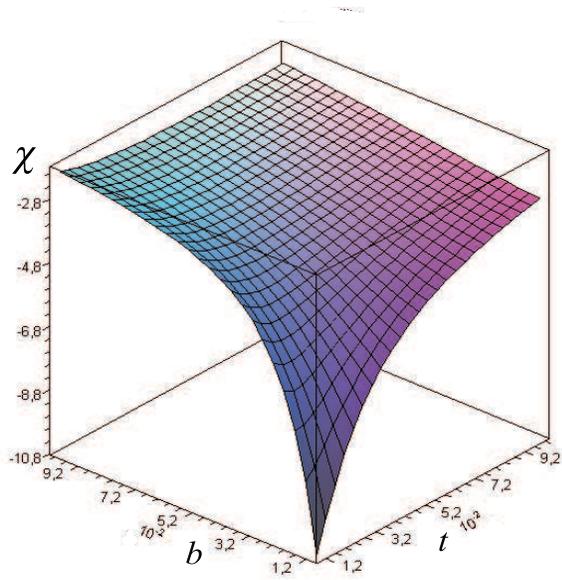}
\caption{\label{Fig2} The susceptibility shape function
$\chi(t,b)$ for q2D systems and variations of $t = (0.01, ..., 0.1)$ and $b = (0.01,...,0.1)$. }
\end{figure}

\begin{figure}
\includegraphics[width=8.5cm, height=8cm, angle=0]{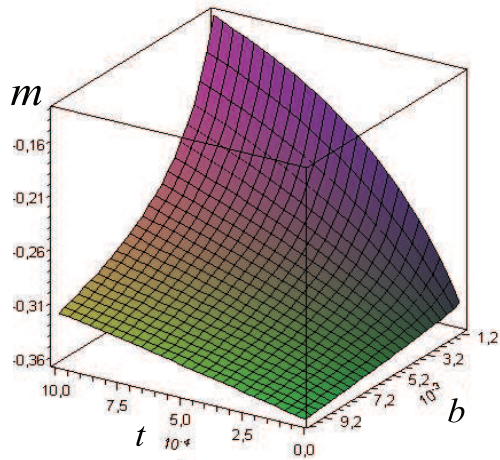}
\caption{\label{Fig3} The magnetization shape function $m(t,b)$
for q2D systems and variations of $t =(0.00001, ..., 0.001)$ and $b = (0.001,...,
0.01)$. }
\end{figure}

\begin{figure}
\includegraphics[width=8.5cm, height=8cm, angle=0]{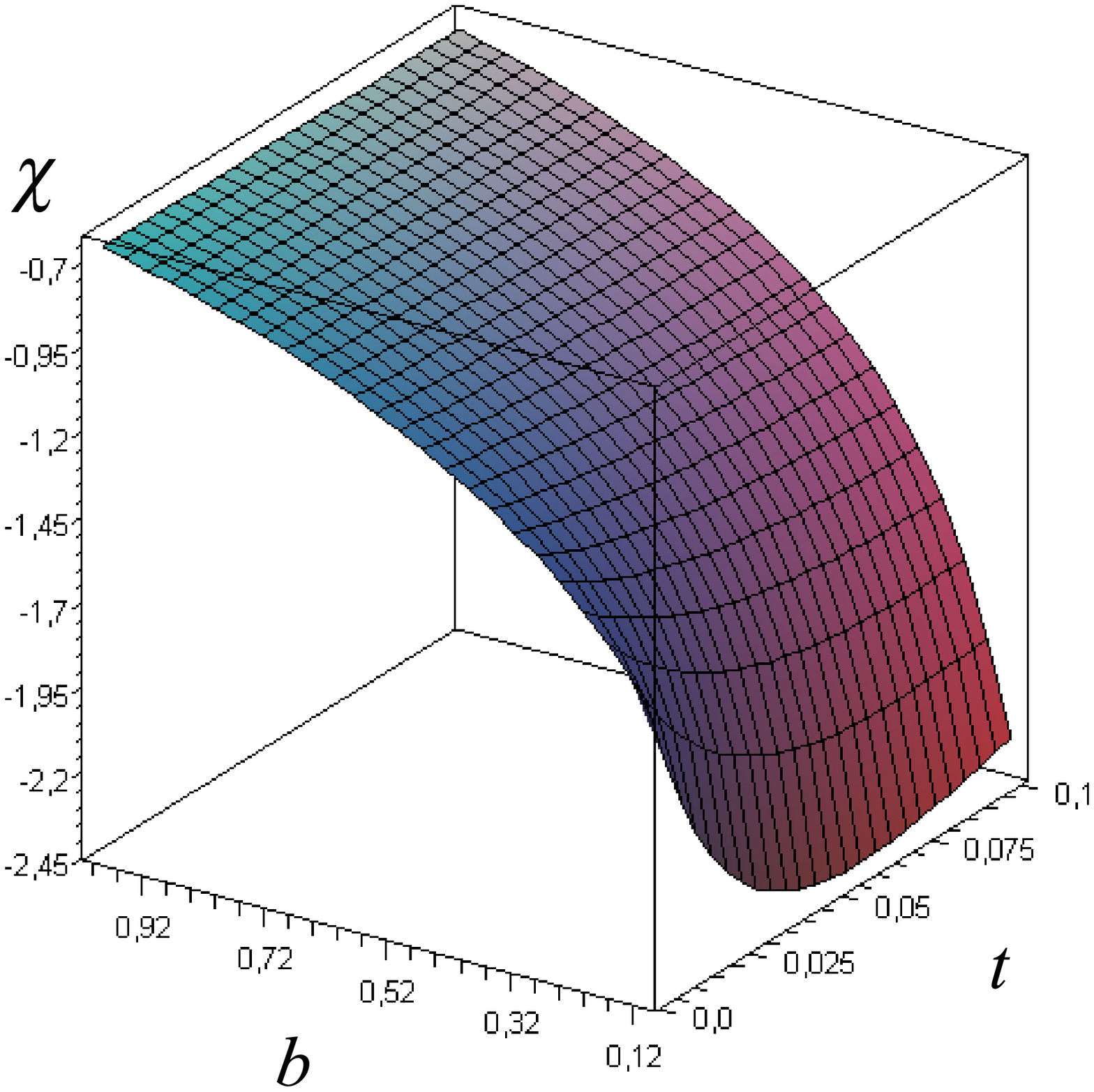}
\caption{\label{Fig4}The susceptibility shape function $\chi
(t,b)$ for q2D systems: $t = (0.001,..., 0.1)$ and $b= (0.1,...,1)$. }
\end{figure}

Having in mind the relations (\ref{Eq26}) and (\ref{Eq27}) as well
as $\partial/ \partial b = B_0\partial /\partial B$, for q2D
systems we obtain $\chi_{dia} = (\rho S/B_0)\chi(t,b)$, where
$\chi(t,b)$ is the susceptibility shape function. The latter is
defined by $\chi = -\partial^2 \tilde{f}(0,t,b)/\partial b^2$ with
$\tilde{f}(t,b) = \tilde{S}(0,t, b)$, where $\tilde{S}(0,t,b)$
denotes $S(0,t,b)$ with $(\pi k_B/\alpha_c) = 1$ for suitable
choice of units, as given by Eq.~(\ref{Eq30}) for $k_z=0$. The
function $\chi (t,b)$ is depicted in Fig.~\ref{Fig2} for $t=
(0.01,..., 0.1)$ and $t= (0.01,..., 0.1)$. As shown in
Fig.~\ref{Fig2}, the shape function $\chi(t,b)$ exhibits a sharp
decrease even at values of $(t,b) \sim (0.01, 0.05$. The minimal
value $\chi \sim -10$ for $t\sim b\sim 0.01$ is a precursor of
divergency, given by Eq.~(\ref{Eq39}). As we see from
Fig.~\ref{Fig2}, the decrease of the function  $\chi(t,b)$ is more
steep along the $t$-axis, and changes its monotonic decrease with
a decrease of value of $t$ at any fixed $b$ changes to an increase
at some finite $t_m(b) >0$, which renders the minimal value of
$\chi$ at given $b$. Obviously, the decrease of $\chi$ is not
symmetric with respect to the axes $t$ and $b$ even in the
pre-critical region $t\sim b\sim 0.01$. The difference in the
behavior of $\chi$ with respect to $t$ and $b$ is better seen in
the strong-$B$ limit ($t \ll b$).

\subsection{Strong-$B$ limit}

For relatively large induction $B$, the leading terms in the sum (\ref{Eq30}) are

\begin{align}
\label{Eq40} S(k_z,t,b) &=  \frac{\ln b}{2} + b\ln\frac{2\alpha_c(1+b)}{\pi
k_B} \nonumber \\ &
+b\ln\left[\frac{\Gamma\left(1/b\right)}{\Gamma\left(1/2b\right)}\right]
+ O(b\varepsilon),
\end{align}
where $b$-independent terms and small term of order $O(bt)$ have
been omitted. This expression of $S(k_z,t,b)$ is valid for any $0
< b \leq 1$ and does not contain $\varepsilon (k_z)$. Therefore,
the result (\ref{Eq40}) can be obtained by setting $t=
(\xi_0k_z)^2 =0$ in Eq.~(\ref{Eq30}) and by applying properties of
the gamma function $\Gamma (z)$~\cite{Abramowitz:1965}.

In this limiting case, $F_{3D}$ is related with $F_{2D}$ by

\begin{equation}
\label{Eq41} F_{3D} = \frac{L_z}{\xi_0}F_{2D},
\end{equation}
as implied by Eq.~(\ref{Eq29}), and $F_{2D}$ is given by

\begin{equation}
\label{Eq42} F_{2D} =   \rho S B_0S(b),
\end{equation}
where $S(b)$ is a short notation of the expression (\ref{Eq40}) of
$S(k_z,t,b) \approx S(0,0,b)$ in the strong-$B$ limit.

The sum (\ref{Eq40}) does not exhibit
any singularity. We shall briefly discuss the case $ \varepsilon \ll b \ll 1$ in
order to reveal and generalize a preceding result for the
diamagnetic moment~\cite{Klemm:1973}; see also Ref.~\cite{Larkin:2005}. For small $b$ we obtain from Eq.~(\ref{Eq40})
that

\begin{equation}
\label{Eq43} S(b) =   b\ln\frac{\sqrt{2}\alpha_c}{\pi k_B} + \frac{11}{12}b^2.
\end{equation}
($b$-independent terms have been once again omitted).

Now one may obtain a simple expressions for the free energies of q2D and
3D superconductors. The q2D free energy $F_{2D}$ will be

\begin{equation}
\label{Eq44} F_{2D} =   \rho S B\left[\ln\frac{\sqrt{2}\alpha_c}{\pi k_B} +
\frac{11B}{12B_0} \right],
\end{equation}

\noindent whereas $F_{3D}$ is given by Eqs.~(\ref{Eq41}) and (\ref{Eq44}).
The diamagnetic moment
${\cal{M}}_{2D}$ and the diamagnetic susceptibility $\chi_{dia}^{(2D)}$ will be

\begin{equation}
\label{Eq45} {\cal{M}}_{2D} =   -\rho S\left[\ln\frac{\sqrt{2}\alpha_c}{\pi k_B} +
\frac{11B}{6 B_0} \right],
\end{equation}
\noindent and

\begin{equation}
\label{Eq46} \chi^{(2D)} =   -\frac{11\rho S }{6B_0},
\end{equation}
\noindent respectively. For 3D systems, in accord with Eq.~(\ref{Eq41}),
${\cal{M}}_{3D} = (L_z/\xi_0){\cal{M}}_{2D}$,
and  $\chi_{dia}^{(3D)} = (L_z/\xi_0)\chi_{dia}^{(2D)}$, where ${\cal{M}}_{2D}$
and $\chi_{dia}^{(2D)}$ are given by
Eqs.~(\ref{Eq45}) and (\ref{Eq46}).

These results are shown in Figs.~\ref{Fig3} and~\ref{Fig4}. For
q2D systems, the magnetization shape function $m = {\cal{M}}/\rho
S $ is given by $m(t,b) = -\partial \tilde{S}(0,t, b)/\partial b$,
where $\tilde{S}(0,t,b)$ is equal to $S(0,t,b)$ for $\pi
k_B/\alpha_c = 1$; see Eq.~(\ref{Eq30}). The function $m(t,b)$ is
shown in Fig.~\ref{Fig3} for $t = (0.0001,...,0.001) \ll b =
(0,001,\dots,0.01)$; q2D systems. When $b$ tends to $0.01$, the
variations of $m(b\sim 0.01, t)$ with $t \in (10^{-5}, 10^{-3})$
are relatively small compared to those for $t\sim b \sim 10^{-3}$.
At given small $t$, $t \sim 10^{-4}$ in Fig.~\ref{Fig3}, $\chi$
slowly increases with the decrease of $b$ following the linear low
(\ref{Eq45}), and tends to $-\ln2/2 \approx 0. 345$ for $b =
10^{-3}$ in accord with Eq.~(\ref{Eq45}); a result, firstly
achieved in Ref.~\cite{Klemm:1973}.

The susceptibility shape function $\chi(t,b)$ of q2D systems is
shown in Fig.~\ref{Fig4} for $t=(0.001,\dots,0.1)$ and $b=
(0.1,\dots,1)$. As seen from Fig.~\ref{Fig4}, the shape function
$\chi (t,b)$ remains finite, provided $b \gg t$ even when $b$
tends to zero. Besides, as we have shown analytically for $b \ll
1$, in the large-$B$ limit this function virtually does not depend
on $t$ for any fixed $0 < b < 1$. In accord with our analytical
result Eq.~(\ref{Eq46}), valid for $t \ll b \ll 1$, Fig.\ref{Fig4}
shows that at fixed $b$, the function $\chi(t,b)$ is almost
constant for variations of $t$ under the condition $t \ll b$. At
fixed $t \ll b$, however, the variations of the function $\chi(b)$
are substantial, in particular, for $b \ll 1$. When $t$ tends to
zero, the function $\chi(t\sim 0, b)$ is bounded from below at
$-11/6$, as seen from both Eq.~(\ref{Eq46}) and Fig.\ref{Fig4}.

\subsection{Discussion of the results: application to $p$-wave ferromagnetic superconductors}

The results for the free energy $F$, the diamagnetic moment
${\cal{M}}$, and the diamagnetic susceptibility $\chi_{dia}$ are
very similar to the respective known results for conventional
non-magnetic superconductors~\cite{Larkin:2005}. In particular we
point out the dependence of these physical quantities on the
parameters  $t$ and $b$, describing the departure of thermodynamic
states from the phase transition line $T_c(P)$. For non-magnetic
superconductors ($\mathbf{M} \equiv 0$), $\mathbf{B} =
\mathbf{H}$, $T_c(M=0) = T_{c0}$ is the usual superconducting
critical temperature at zero external magnetic field, and we
reveal the known results for 3D and q2D standard superconductors,
summarized in the review~\cite{Larkin:2005}.

In $p$-wave ferromagnetic superconductors, the magnetization
$\mathbf{M}$ in zero external magnetic field $\mathbf{H}$ is
different from zero along the whole phase transition line $T_c(P)$
and the shape of critical temperature $T_c(P)$ in zero external
magnetic field $\mathbf{H}$ strongly depends on the magnetization
$\mathbf{M}$, as given by Eq.~(\ref{Eq25}). Thus, except for a
very narrow domain of the $T-P$ phase diagram above the critical
pressure $P_c$, the magnetization above the line $T_c(P)$ is
always large and, hence, for this case, we should consider large
values of the induction $B$ even when the external magnetic field
$\mathbf{H}$ is small or equal to zero. The important quantity in
our consideration is the induction $B$, because the latter enters
in the magnetic frequency $\omega_B$ and the magnetic length
$a_B$. Now the magnetic induction $B$ plays a role similar to that
of external magnetic field $\mathbf{H}$ in the theory of
diamagnetic moment and diamagnetic susceptibility in usual
superconductors~\cite{Larkin:2005, Pitaevskii:1980}. Therefore,
for $p$-wave ferromagnetic superconductors with phase diagrams of
the types shown in Fig.\ref{Fig1}, we should use only those of our
results,  which correspond to the large-$B$ limit. Our results
show that both diamagnetic moment and diamagnetic susceptibility
do not exhibit any singularity and remain finite up to $T=T_c(P)$
along the most part of the FM-FS phase transition line $T_c(P)$.
This means that the diamagnetic singularities are dumped by the
ferromagnetic order.

The results in the weak-$B$ limit could be valid in a close
vicinity of the critical pressure $P_c$, where the magnetization
$M$ of the ferromagnetic phase is very small and the criterion for
weak-$B$ limit is fulfilled: $(H + 4\pi M) \ll B_0$. Then the
results in this limit will be valid for enough small external
field $H$ and $M\simeq M_c = M(T_c) \ll B_0$. In this case, as
mentioned in Sec.~IV.A.1, the diamagnetic moment ${\cal{M}} < 0$
will exist even for $H=0$ and will be proportional to the
ferromagnetic moment $M \simeq M_c$; see Eq.~(\ref{Eq35}).
According to Eq.~(\ref{Eq35}), the 3D superconductor will have a
negative total magnetic momentum $M_{tot} = {\cal{M}} + VM$ at
$H=0$ provided

\begin{subequations}
\begin{equation}
\label{Eq47a} \frac{\pi\rho}{3B_0\xi_0}>t^{1/2}.
\end{equation}

According to Eq.~(\ref{Eq38}), the total magnetic moment in 2D superconductors will be negative,
$M_{tot} = {\cal{M}} + SM < 0$, provided

\begin{equation}
\label{Eq47b} \frac{2\pi\rho}{3B_0}>t.
\end{equation}

\end{subequations}

When the criteria (\ref{Eq47a}) and (\ref{Eq47b}) are satisfied
the diamagnetism prevails and the overall magnetization of the
system is negative. However, under certain conditions, some
relevant fluctuation contribution of the magnetization vector
$\mathbf{M}$ may occur, and this is an issue which need a study
beyond the Gaussian approximation. A reliable application of our
results in the weak-$B$ limit could be performed in ferromagnetic
superconductors, where a line of phase transition of type N-FS
exists, i.e., when the lines of the N-FM and FM-FS phase
transitions of first order meet at some finite temperature
critical-end point ($T, P_c^{\prime} \sim P_c$) which is connected
with the zero temperature point $(0,P_c)$ by a second-order (N-FS)
phase transition line (Sec.I).

\section{Conclusion}

Introducing an advanced theoretical approach, we have been able to investigate
 the basic properties of the superconducting
fluctuations in $p$-wave ferromagnetic superconductors in zero
external magnetic field. For the presence of a strong
magnetization due to the ferromagnetic state, the $p$-wave
ferromagnetic superconductors with $T_F(P) \gg T_c(P)$ exhibit a
diamagnetic behavior, which is typical for usual superconductors
in the strong-$H$ limit. For this type of ferromagnetic
superconductors we have demonstrated a form of universality. It is
known~\cite{Larkin:2005} that at a strong external field
$\mathbf{H}$, the diamagnetic quantities do not exhibit
singularities. Here the same quantities undergo the same dumping,
i.e. lack of singularities in the strong-$B$ limit, including
their values at external field equal to zero.

In the weak-$B$ limit the diamagnetic moment and the diamagnetic
susceptibility exhibit scaling singularities with respect to the
parameter $t$ of type known from the theory of non-magnetic
superconductors~\cite{Belich:2010}. As demonstrated in
Ref.~\cite{Belich:2010} for 3D geometry, and here for both 2D and
3D superconductors, the scaling amplitudes for $p$-wave
ferromagnetic superconductors are quite different from the known
scaling amplitudes for nonmagnetic
superconductors~~\cite{Larkin:2005}. The difference is due to the
existence of ferromagnetic moment $M$ above the FM-FS phase
transition line $T_{FS}(P)$; see Fig.~\ref{Fig1}. Therefore, our
new results in the weak-$B$ limit may have application to $p$-wave
superconductors with $T-P$ diagrams of shape shown in
Fig.~\ref{Fig1}, where the FM-FS phase transition line lies below
the N-FM phase transition line. Besides, in order to apply the
weak-$B$ limit results, the ferromagnetic moment $M$ should be
enough small. Thus, the results for weak-$B$ could be applied only
in a close vicinity of the critical pressure $P_c$, where the
N-FM and FM-FS phase transition lines are very close to each other
and the ferromagnetic states between them possess a small
magnetization $M$. If the N-FM phase transition in this domain of
$T-P$ diagram is of first order, as indicated by the experimental
data, the ferromagnetic fluctuations are suppressed and could not
affect on the fluctuation diamagnetism.

The weak-$B$ limit may be applied to $p$-wave superconductors
containing a N-FS line of phase transition. Then the fluctuation
diamagnetism in the $N$-phase will be described by the known
formulae~\cite{Larkin:2005}. For the lack of ferromagnetic moment
in the N-phase ($M = 0, B=H$), in such cases we should consider
''weak-$H$ limit~\cite{Larkin:2005}. The physics of $p$-wave
ferromagnetic superconductors is not limited to the ferromagnetic
compounds enumerated in this paper and those discovered until now.
In future, new substances exhibiting $p$-wave ferromagnetic
superconductivity with different shape of the $T-P$ phase diagram
may be discovered. The theory predicts a variety of possible $T-P$
phase diagrams, including both cases with $T_F > T_s$ and $T_s >
T_F$. In the last case, stable pure (non-magnetic) phases are
possible~\cite{Shopova:2003}. This means that the results for the
weak-$H$ may have a wider application.

Except for the location of the phase transition line $T_c(P)$ at $\mathbf{H}=0$,
given by Eq.~(\ref{Eq25}), all results for the diamagnetic quantities $p$-wave
superconductors in Gaussian approximation can be obtained from the known results
for usual (non-magnetic) superconductors by the substitution $\mathbf{H} \rightarrow \mathbf{B}$.
This is a form of universality, deduced in the present paper.

We have used the Gaussian approximation, which is not valid in the
critical region~\cite{Uzunov:2010} of anomalous fluctuations. As
the critical region of real ferromagnetic superconductors with
spin-triplet electron pairing is often very narrow and, hence,
virtually of no interest, the present results can be reliably used
in interpretation of experimental data for real itinerant
ferromagnets, which exhibit low-temperature $p$-wave
superconductivity triggered by the ferromagnetic order.

\end{document}